\documentclass{desyproc}

\begin{document}
\title{Physics of tau and charm}

\author{{\slshape Yifang Wang\thanks{yfwang@ihep.ac.cn} }\\[1ex]
Institute of High Energy Physics, Beijing 100049 }

\contribID{xy}

\confID{800}  
\desyproc{DESY-PROC-2009-xx}
\acronym{LP09} 
\doi  

\maketitle

\begin{abstract}
The physics of tau and charm started in early 70's after J/$\psi$
and $\tau$ were discovered. Since then several dedicated accelerators
and experiments were built with increasing luminosities and studies
on light hadron spectroscopy, charmonium, electroweak and QCD were
never interupted. New interests and surprises are not rare in this area.
With the newly built BEPCII/BESIII, an even brighter future is foreseen.
\end{abstract}

\section{introduction: physics at tau-charm colliders}

Since the successful test of ADA, several electron-positron
colliders were built in late 60's and early 70's. The most
successful one is SPEAR, which discovered both the $\psi$ and tau,
marking the beginning of the tau-charm physics. Many members of
the charmonium family and charmed mesons were then discovered at
SPEAR and Doris, and many unknowns and controversies were
resolved up to 80's.

\begin{figure}[hb]
\centering
\includegraphics*[width=85mm]{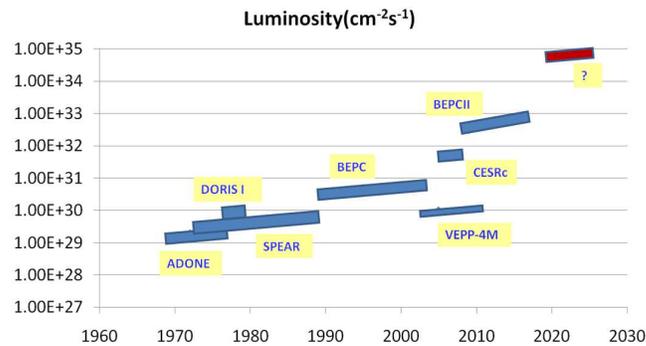}
\caption{A history of accelerators running for tau-charm
physics.}\label{fig:1}
\end{figure}

The precision physics at tau-charm energy region began from
BEPC/BES, the first accelerator even built for high energy physics
in China in early 90's. The luminosity were improved by an order
of magnitude over SPEAR, and this record was surpassed by
CESR-c in 2004, as shown in Fig.~\ref{fig:1}.

The newly built BEPCII/BESIII is an upgrade to the previous
BEPC/BES~\cite{talk,besiii}. The designed peak luminosity is $1\times
10^{33}$ cm$^{-2}$s$^{-1}$, another order of magnitude over
CESR-c. The race may continue with the idea of a super-tau-charm
factory with a luminosity of about $1\times 10^{35}
cm^{-2}s^{-1}$.

Why we are so interested in building tau-charm colliders in the
last 40 years, even now in the era of LHC, and possibly in the
future? In fact, J/$\psi$ and its family can be produced at
$e^+e^-$ colliders with huge cross sections and abundant
resonances, providing an ideal laboratory for charm, charmonium
and QCD studies. Charm quark is actually a bridge between pQCD and
non-pQCD, and relevant information becomes a ruler for Lattice
QCD. Charmonium decays through the so-called three-gluon loop is
one of the best channels to search for glueballs and hybrids. The
threshold production of charmonia and taus has a lot of
advantages on background suppression, kinematic constraints and quantum correlations.
In the era of LHC, high precision flavor physics is complementary
since new phenomena at high energies should also evident via
virtual loops and secondary effects at lower energies.

The latest progress of tau-charm physics and prospects at the
newly built high luminosity tau-charm collider, BEPCII, is
summarized in the book "Physics at BESIII~\cite{ref:yellowbook},
which covers all the areas including the charm physics, charmonium
physics, tau physics, QCD studies and light hadron spectroscopy.
Examples of highlights include $D\bar{D}$ mixing, precision
measurement of CKM matrix elements and the tau mass, charmonium
transition and spectroscopy, exotic hadron searches, new hadrons
above the open charm threshold, etc.

In this talk, I will select a few topics to report the progress in
this field.

\section{CLEOc: a fruitful short program on charm physics}

CESRc started its charm program since 2004 and ceased operation in
2008. Although very short, it is a very fruitful program in charm and
charmonium physics.

Threshold production of charmed mesons is of particular importance
since D and $\bar D$ are doubly produced at $\psi(3770)$ at rest.
$D\bar D$ mixing can then be studied in an almost background-free
environment with quantum correlations. Even the statistics is low and there is no time-development, the double-tag technique allows to reduce systematic errors, hence a complementary to that at B-factories. CLEO
reported a first determination of cos$\delta$ using the quantum
correlation between two D's produced at rest from $\psi(3770)$
decays~\cite{ref:CLEOcosd}. In fact, due to the mixing, tagging
one $D^0$ in a CP eigenstate, the other side is a mixture of $D^0$
and $\bar D^0$ with an event rate proportional to $B_1 B_2(1+2rcos
\delta)$, where $B_1$ and $B_2$ are branching ratios of D's at
each side, and cos$\delta$ is the quantum correlation, which is
related to the mixing parameters. A global fit is performed for 8
hadronic D decay channels and $\delta$ is determined to be
$(22^{+11+9}_{-12-11})^o$, limited by statistics. Clearly with
BESIII, a significant improvement can be expected.

$D\bar D$ mixing has been firmly established, thanks mainly to
B-factories with great statistical advantages. A global fit shows
that mixing is established at 10.2 $\sigma$ level, and consistent
with CP conservation~\cite{ref:hfag}. These results are consistent
with the Standard Model as well as many New Physics models. In
fact, Standard Model can not give a reliable prediction due to the
complication at hadron level. In order to understand the origin of
the mixing, we need to integrate all the flavor physics results,
correlate them with other mixing results, have more data on rare
decays and CP violation limits to constrain New Physics models.
CLEO searched for the CP violation in many D and $D_s$
decays~\cite{ref:CLEOCP}. BESIII at a high luminosity machine can
in fact improve the limit significantly.

CKM matrix elements can be measured precisely via leptonic and
semi-leptonic decays of D mesons. In the leptonic case,
the decay width follows
$$\Gamma(D_{(s)}\rightarrow l\nu) = f^2_{D_{(s)}}\vert V_{cq}\vert ^2{G^2_F\over {8\pi}}
m_{D_{(s)}}m^2_l(1-{m^2_l\over m^2_{D_{(s)}}})^2 .$$
CLEO recently
reported their measurements of $D^+\rightarrow \mu^+\nu $, $D^+_s
\rightarrow \mu^+\nu $ and $D^+_s \rightarrow\tau^+\nu $, giving
$f_D = (205.8 \pm 8.5\pm 2.5)$ MeV, $f_{D_s} =(259.5 \pm 6.6 \pm
3.1)$ MeV~\cite{ref:CLEOfd}. While $f_D$ is in perfect agreement
with the prediction of lattice QCD~\cite{ref:QCDpred}, i.e.
$(208\pm 4)$ MeV from the HPQCD-UKQCD group,
$f_{D_s}$ is 2.3 $\sigma$ away from the prediction of $(241\pm 3)$ MeV.
BESIII may resolve this issue with a larger statistics and better precision.

\begin{wraptable}{r}{0.5\textwidth}
\centerline{\footnotesize\begin{tabular}{|l|r|} \hline decay mode
& branching fraction  \\ \hline
$D^+\rightarrow \eta e^+\nu $  & $0.133\pm 0.020 \pm 0.006$  \\
$D^+\rightarrow \eta' e^+\nu $ &   $<$0.035  \\
$D^+\rightarrow \phi e^+\nu $  &   $<$ 0.016  \\  \hline
$D^+_s\rightarrow \eta e^+\nu $ & $2.48\pm 0.29\pm 0.13$     \\
$D^+_s\rightarrow \eta' e^+\nu $ & $0.91\pm 0.33\pm 0.05$      \\
$D^+_s\rightarrow \phi e^+\nu $ & $2.29\pm 0.37\pm 0.11$      \\
$D^+_s\rightarrow K^0 e^+\nu $ & $0.37\pm 0.10\pm 0.02$     \\
$D^+_s\rightarrow K^{*0} e^+\nu $ & $0.18\pm 0.07\pm 0.01$      \\
$D^+_s\rightarrow f_0(\rightarrow \pi^+\pi^-) e^+\nu $ & $0.13\pm
0.04\pm 0.01$      \\\hline
\end{tabular}}
\caption{Branching ratios of several new semi-leptonic decay modes
measured by CLEO.} \label{tab:semi}
\end{wraptable}

\begin{figure}[hb]
\centering
\includegraphics*[width=75mm]{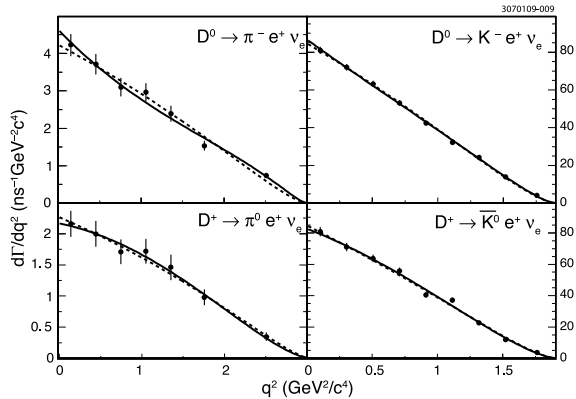}
\caption{The momentum spectrum of semi-leptonic decays from CLEO
experiment} \label{fig:semi}
\end{figure}

In the semi-leptonic case, the differential decay width follows
$${d\Gamma(X\rightarrow X'l\nu)\over dq^2} = [f_+^{X\rightarrow X'}(q^2)\vert V_{Qq} \vert]^2
{G^2_F \over 24\pi^3}p^3_{X'}.$$
By fitting this formula with data
and using LQCD prediction of $f_+^K(0)$ and $f_+^{\pi}(0)$, as
shown in Fig.~\ref{fig:semi}, CLEO obtained new CKM matrix element
measurement~\cite{ref:semi}, $|V_{cd}|=0.234\pm 0.007(stat.)\pm
0.002(syst.)\pm 0.025(LQCD)$, and $|V_{cs}|=0.985\pm 0.009(stat.)
\pm 0.006(syst.)\pm 0.103(LQCD) $. Several new D and $D_s$ semi-leptonic
decay modes are observed for the first time by
CLEO~\cite{ref:seminew}, as listed in Table~\ref{tab:semi}. These new
D decay modes are interesting for glueball searches, while new
$D_s$ decay modes are Cabibbo-suppressed and scalers.

Only a small fraction of charm physics results from CLEO are reported here.
For more information, please refer to recent publications of the CLEO collaboration. Now let's turn to charmonium physics.

Charmonium family, an interesting lab for pQCD and non-pQCD, can
be used to calibrate LQCD. Their productions, transitions, decays
and the spectroscopy are not fully understood yet and examples of
interesting and long-standing issues including the $\rho\pi$
puzzle, mixing state, missing states, and new XYZ states. For detailed
discussion, please refer to reference~\cite{ref:yellowbook}.

\begin{figure}[htb]
\centering
\includegraphics*[width=55mm]{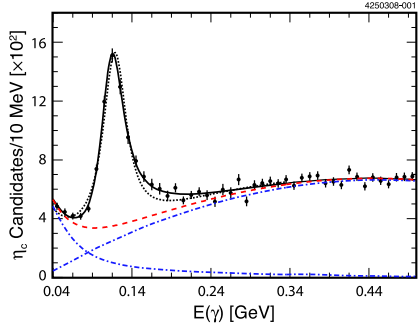}
\caption{ Fits to the photon spectrum in exclusive $J/\psi
\rightarrow \gamma \eta_c$ decays using relativistic Breit-Wigner
(dotted) and modified (solid) signal line shapes convolved with a
4.8 MeV wide resolution function. } \label{fig:etac:cleo}
\end{figure}

$\eta_c$ is the lowest state of the charmonium family, its mass
and width are hence critical. However, current mass measurements
are not consistent, and this problem is traced by CLEO to be the
distorted line-shape of $\eta_c$~\cite{ref:etac} from a standard
Breit-Wigner form,as shown in
Fig~\ref{fig:etac:cleo}. The reason is not known yet, and CLEO fitted the data with a modified empirical Breit-Wigner formula. We are
waiting for results from BESIII on exclusive channels and energy
dependent $\psi(1S, 2S)\rightarrow \gamma\eta_c$ matrix element
measurements.

\begin{figure}[htb]
\centering
\includegraphics*[width=50mm]{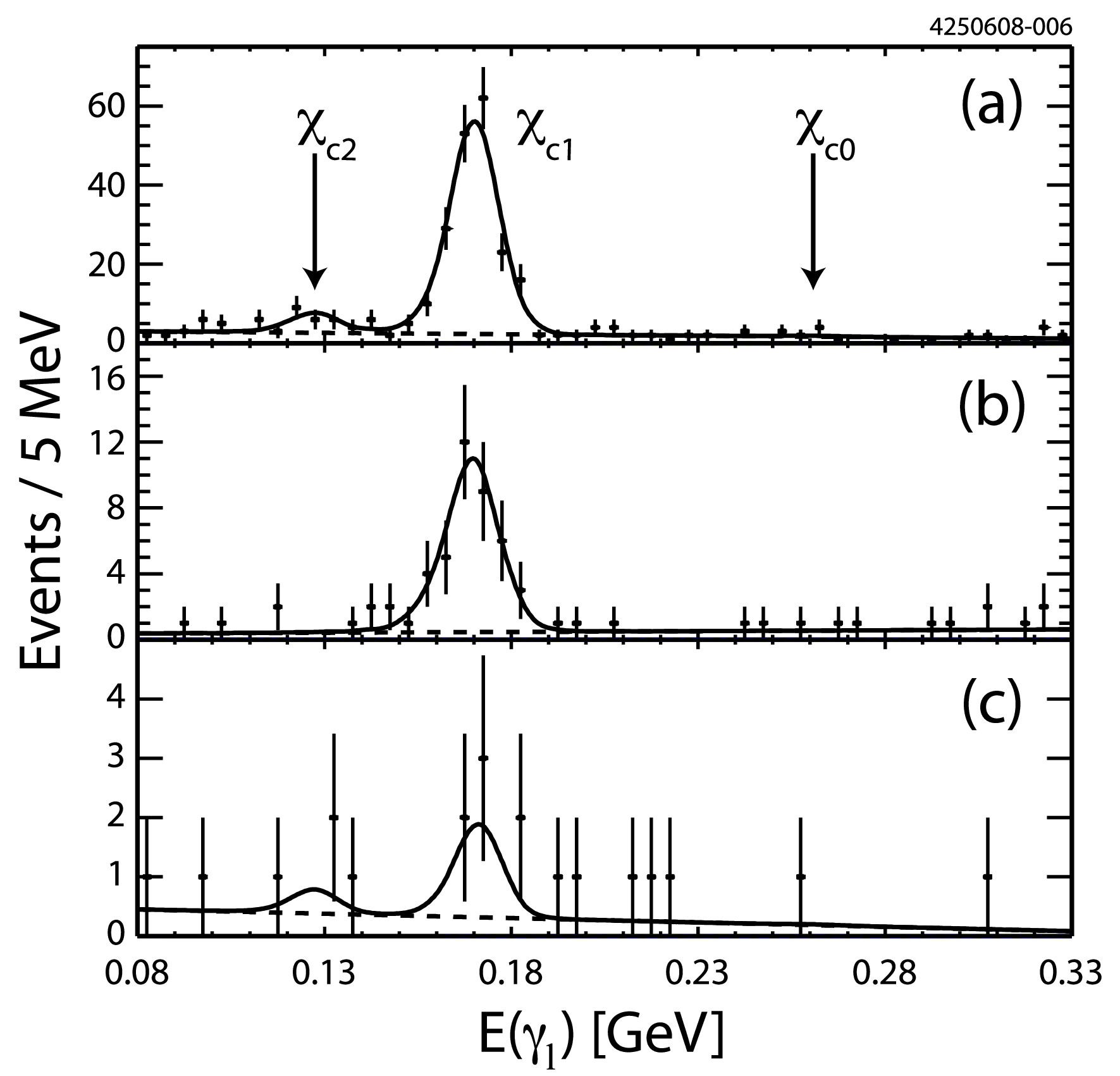}
\caption{Observed signal of $\chi_{cJ}\rightarrow \gamma V$. The
$\psi(2S)\to\gamma\chi_{cJ}$ transition photon
($\gamma_\mathrm{l}$) energy distribution for (a)
$\chi_{cJ}\to\gamma\rho^0$, (b) $\chi_{cJ}\to\gamma\omega$, and
(c) $\chi_{cJ}\to\gamma\phi$ candidates.  The data are shown by
the points; the fit is shown as a solid line.  The background
component of the fit is indicated by the dashed line.}
\label{fig:CLEOgv}
\end{figure}

$\chi_{cJ}$ from $\psi'$ decays is ideal for the light hadron
spectroscopy since they have clean and multiple $J^{PC}$
states. In two body decays they can be used to study the
role of the color octet mechanism and to probe the gluon content in
final states. CLEO reported results for two-baryon final states
previously~\cite{ref:CLEOgbb}, and two-meson final states recently
including $\pi\pi$, $\eta\eta$, $\eta\eta'$, $\eta'\eta'$, $KK$,
etc.~\cite{ref:CLEOgpp}. Radiative decay processes of $\chi_{cJ}$
to light vectors, $\chi_{cJ}\rightarrow \gamma(\rho,\omega,\phi)$,
similar to that of the glueball production of $J/\psi\rightarrow \gamma
f_J$, have been searched for. For the first time, the decay
modes of $\chi_{c1} \rightarrow \gamma\rho$ and $\chi_{c1}
\rightarrow \gamma\omega $ are observed~\cite{ref:CLEOgv}.
Figure~\ref{fig:CLEOgv} shows the observed signals and
table~\ref{tab:CLEOgv} list the results in comparison with the
prediction based on pQCD calculations~\cite{ref:gvpred}. However,
the prediction is one order of magnitude below the observation.

\begin{table}[htp]
\begin{center}
\footnotesize
\begin{tabular}{|c|c|c|c|}\hline
decay mode & BR$\times 10^6$ & U.L.[$10^{-6}$] & pQCD[$10^{-6}$]\\
\hline
$\chi_{c0}\rightarrow \gamma\rho^0$ &                    &    $<$9.6  & 1.2  \\
$\chi_{c1}\rightarrow \gamma\rho^0$ & $243\pm 19\pm 22$   &          & 14  \\
$\chi_{c2}\rightarrow \gamma\rho^0$ & $25\pm 10^{+8}_{-14}$     &    $<$50  & 4.4  \\
$\chi_{c0}\rightarrow \gamma\omega$ &                    &    $<$8.8  & 0.13  \\
$\chi_{c1}\rightarrow \gamma\omega$ & $83\pm 15 \pm 12$  &          & 1.6  \\
$\chi_{c2}\rightarrow \gamma\omega$ &                    &    $<$7.0  & 0.5  \\
$\chi_{c0}\rightarrow \gamma\phi  $ &                    &    $<$6.4  & 0.46  \\
$\chi_{c1}\rightarrow \gamma\phi  $ &$12.8\pm 7.6\pm 1.5$&    $<$26  & 3.6  \\
$\chi_{c2}\rightarrow \gamma\phi  $ &                    &
$<$13 & 1.1  \\ \hline
\end{tabular}
\caption{Measured Branching ratios or up limits in comparison with
pQCD calculations} \label{tab:CLEOgv}
\end{center}
\end{table}

CLEO also reported charmonia($J/\psi$, $\psi(2S)$ and
$\psi(3770)$) radiative decays to pesude-vectors, including
$\pi^0$, $\eta$ and $\eta^{\prime}$~\cite{ref:CLEOgpi}.
Improvements over previous measurements on $J/\psi$ and $\psi(2S)$
decays were observed, while no $\psi(3770)$ decays was observed. A
new decay mode of $J/\psi \rightarrow \gamma\gamma\gamma$ was
observed~\cite{ref:CLEOggg}, and the branching fraction is
measured to be $(1.2\pm 0.3\pm 0.2)\times 10^{-5}$. This is the
quarkonium analogue of ortho-positronium decay, and no similar
decays have been observed for any particles so far.

The last member of the charmonium family under the open charm
threshold, $h_c$ was discovered by CLEO~\cite{ref:CLEOhc1} and an
updated product branching fraction, $B(\psi(2s)\rightarrow \pi^0
h_c)\times B(h_c\rightarrow \gamma\eta_c)$, was reported to be
$(4.19\pm 0.32 \pm 0.45) \times 10^{-4}$~\cite{ref:CLEOhc},
averaging the inclusive and exclusive channels. Later I will
report the first BESIII results which improves this
measurements. In fact, with a much larger data sample and a great
detector, BESIII will improve significantly all the results
mentioned above, and new discoveries are expected.

\section{KEDR: a special dedication to mass measurement}
\begin{wraptable}{r}{0.5\textwidth}
\centerline{\footnotesize\begin{tabular}{|l|r|}  \hline particles
& mass(MeV)
\\ \hline
tau &   $1776.69^{+0.17}_{-0.19} \pm 0.15$     \\
J/$\psi$ &  $3096.924 \pm 0.010 \pm 0.017$ \\
$\psi(2s)$ &  $3686.125 \pm 0.010 \pm 0.015$ \\
$\psi(3770)$ & $3772.8 \pm 0.5 \pm 0.6 $ \\
$D^{\pm}$ & $1869.32 \pm 0.48 \pm 0.21 $  \\
$D^0$ & $1865.53 \pm 0.39 \pm 0.24 $  \\\hline
\end{tabular}}
\caption{Recent mass measurements at KEDR} \label{tab:kedr}
\end{wraptable}
The VEPP-4M accelerator and the KEDR detector at Novosibirsk in
Russia started operation in 2002 and the luminosity is about
$1\times 10^{30} cm^{-2}s^{-1}$. A special physics program was
performed by calibrating the beam energy precisely. Two
techniques are developed: Resonance Spin Depolarization
with a precision better than 30 keV and Back
Compton Scattering with a precision better than 150 keV. The mass
of particles in the tau-charm energy region, including tau,
J/$\psi$,$\psi'$, D mesons etc. are measured to an un-precedent
precision. Table~\ref{tab:kedr} lists their
results\cite{ref:mass1}. Results are consistent with previous
measurements and further improvements are expected. The Back
Compton Scattering technique will be used at BESIII and an even
more precised tau mass measurement is expected.

\section{BESII: a final legacy}

The partial upgrade of the BES detector, called BESII, stop
operation in 2004, however, physics results on light hadron
spectroscopy and QCD studies are still coming based on the
existing data sample. Since the production cross section of
J/$\psi$ is huge, its decay is an ideal place for light hadron
spectroscopy study. A few examples are given here.

\begin{figure}[htb]
\vspace{-0.5cm}
\centerline{\includegraphics[width=0.40\textwidth]{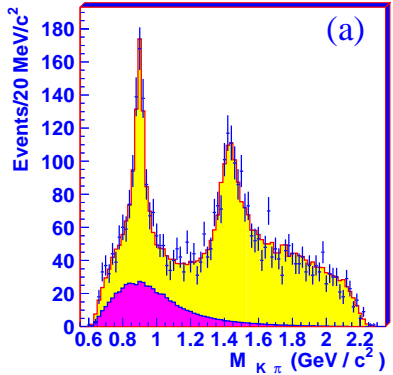}}
\caption{The observed $\kappa^{\pm}$ signal at BESIII from an
invariant mass spectrum of $K^{\pm}\pi^0$. Shaded area are $\kappa$ signals.} \label{fig:kappa}
\end{figure}

$\kappa$ is a very interesting particle needed by the Chiral
Perturbative Theory. There was a hot debate since it was observed
for the first time in $K\pi$ scattering. The E791 experiment found
the evidence of neutral $\kappa$ in 2004 from $D^+\rightarrow
K^-\pi^+\pi^+$ and BESII firmly established its existence in 2006
from $J\psi\rightarrow K^{*0}K\pi\rightarrow K\pi K\pi$
decays~\cite{ref:kappabes2}. CLEO reported the necessity of
charged $\kappa$ in $D^0 \rightarrow K^+K^-\pi^0$, however, no
$\kappa^\pm \rightarrow K^\pm\pi^0$ is needed in BABAR data. BESII
recently observed charged $\kappa$ in $J/\psi \rightarrow
K^{*\pm}\kappa^{\mp}\rightarrow K_s\pi^{\pm}K^{\mp}\pi^0$, as
shown in Fig.~\ref{fig:kappa}~\cite{ref:kappabes3}. The pole
position is measured to be $(841\pm 51^{+14}_{-28})-i(288\pm 101^{
+64}_{-30}) MeV/c^2$, in consistent with that of the neutral one.

In addition to light hadron physics, QCD studies at the tau-charm
energy region is of particular importance since it is at the
boundary between pQCD and non-pQCD. Precision measurement of the
R-value in this region will provide valuable input for vacuum
polarization, improving the prediction of Higgs mass and g-2.
BESII recently reported a new measurement of R at the
center-of-mass energies of 2.6 GeV, 3.07 GeV and 3.65 GeV
respectively, reducing the error from about 6\% to
3.5\%~\cite{ref:rvalue}.

\section{BESIII: a bright future}

The newly completed upgrade of Beijing Electron-Positron
Collider(BEPCII) and the new detector(BESIII) represents the
future of the field~\cite{talk}. BEPCII is a double-ring accelerator with a designed peak luminosity of $10^{33}$ cm$^{-2}s^{-1}$ at
a beam current of 0.93 A. Both the machine and the detector worked
remarkably well since beginning and world largest data samples of
J/$\psi$ and $\psi'$ have been collected. It is
believed that physics at the tau-charm region will be renewed
dramatically and important discoveries will be possible.In the
following I will gave a short summary about their performance and
the initial results recently published.

\subsection{Status of BEPCII/BESIII and data taking}

The BEPCII/BESIII upgrade started in 2003 and successfully completed in 2008. BEPCII managed to accumulate a beam current of 500 mA in the storage ring, and obtained a collision luminosity close to $10^{32}$cm$^{-2}$s$^{-1}$ in March 2008. While the BESIII detector completed installation at the end of 2007 and the first full cosmic-ray event was recorded in March 2008. The detector was successfully moved to the interaction point on April 30, 2008.
With a careful tuning of the machine, the first $e^+e^-$ collision event
was recorded by the BESIII detector on July 19, 2008, and a total
of 14 million $\psi'$ events was collected until Nov. 2008. Over
this period, the BEPCII performance continued to improve by the
lattice optimization, system debugging, and vacuum improvements.
After a 1.5-month synchrotron radiation run and a winter
maintenance, the machine resumed collision and its luminosity
gradually improved from $1\times 10^{32}$ cm$^{-2}s^{-1}$ to
$3\times 10^{32}$ cm$^{-2}s^{-1}$.

Starting from March of 2009, BES-III successfully collected 100
million $\psi(2S)$ events and 200 million $J/\psi$ events, about a
factor of 4 larger than the previous data samples from CLEO-c and
BES-II, respectively. The peak luminosity was stable, typically at
the level of $2\times 10^{32}$ cm$^{-2}$s$^{-1}$ during the data
taking at $\psi(2S)$, and $0.6\times 10^{32}$ cm$^{-2}$s$^{-1}$ at
J/$\psi$. An energy scan of the $\psi(2S)$ line-shape shows that
the beam energy spread is about 1.4 MeV, and the effective peak
cross section of $\psi(2s)$ is about 700 nb.
The data taking efficiency of the detector is
more than 85\%.

\begin{figure}[htb]
\centering\vspace{-0.5cm}
\includegraphics*[width=80mm,angle=90]{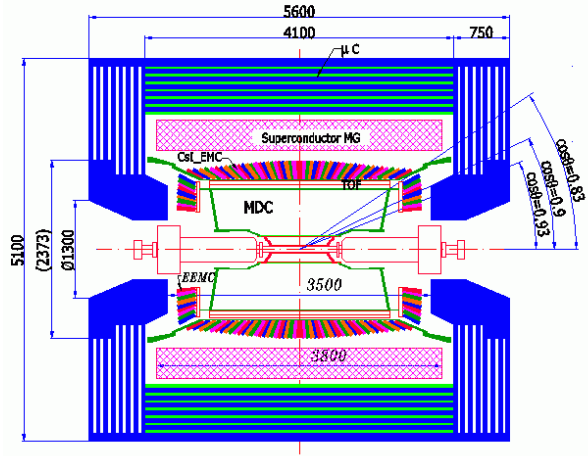}
\caption{A schematic view of the BESIII detector.}
\label{fig3:besiii}
\end{figure}

The BESIII detector~\cite{talk,besiii}, as shown in
Fig.~\ref{fig3:besiii}, consists of the following main components:
1) a main draft chamber (MDC) equipped with about 6500 signal
wires and 23000 field wires arranged as small cells with 43
layers. The designed single wire resolution is 130 $\mu m$ and the
momentum resolution 0.5\% at 1 GeV; 2) an electromagnetic
calorimeter(EMC) made of 6240 CsI(Tl) crystals. The designed
energy resolution is 2.5\%@1.0 GeV and position resolution 6mm@1.0
GeV; 3) a particle identification system using Time-Of-Flight
counters made of 2.4 m long plastic scintillators. The designed
resolution is 80 ps for two layers, corresponding to a K/$\pi$
separation (2$\sigma$ level) up to 0.8 GeV; 4)a superconducting
magnet with a field of 1 tesla; 5) a muon chamber system made of
Resistive Plate Chambers(RPC).

\begin{figure}[htbp]
\centering\vspace{-0.5cm}
\includegraphics*[width=120mm]{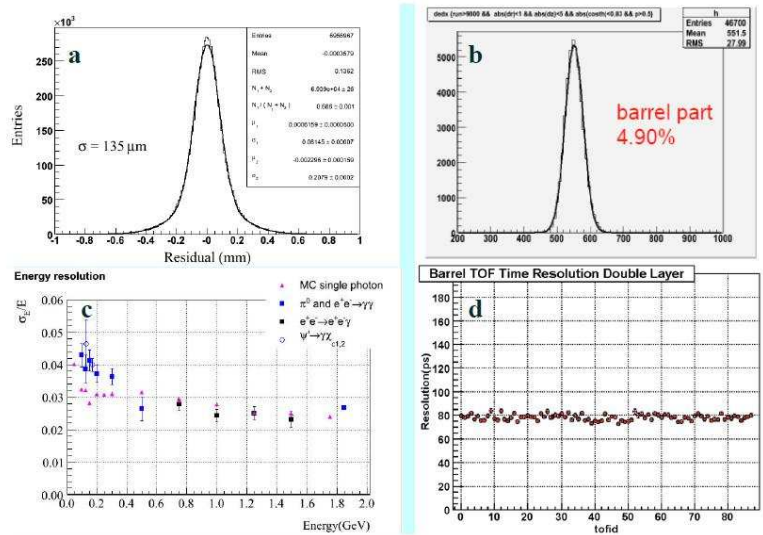}
\caption{Main performance parameters of the calibrated BESIII
detector: a) Single wire resolution of the drift chamber; b) dE/dx
resolution of the drift chamber in the barrel part(w/ all wires);
c) energy resolution of the CsI(Tl) crystal calorimeter as a
function of photon energy from different physics processes; d)
time resolution of TOF counters averaged over two layers for each
counter ID in phi direction.} \label{fig4}
\end{figure}

The detector was calibrated using the $\psi(2S)$ events and the
main performance parameters of the BES-III detector is shown in
Fig.~\ref{fig4}. Clearly, the detector is in a very good condition
and all the design specifications have been satisfied.

A comprehensive Monte Carlo simulation code, largely based on the
first principle of particles interacting with detector materials,
was developed to model the performance of the
BES-III detector. A good agreement was observed, not only
on average numbers, but also on the details functional shape. This
agreement ensures the well control of systematic errors and
precision physics measurement.

\subsection{Preliminary physics results}

Physics at BESIII are very rich~\cite{ref:yellowbook}. An initial
physics program has been planned for the $\psi(2S)$ data set,
including, but not limited to, the following topics:
\begin{itemize}
\item  Spin-singlet studies($h_c$, $\eta_c$, $\eta^\prime_c$);
\item $\psi(2S)$ hadronic decays ($\rho\pi$ puzzle, new states);
\item $\chi_c$ decays (search for new states and new decays).
\end{itemize}

\begin{figure}[htbp]
\centering\vspace{-0.5cm}
\includegraphics*[width=80mm]{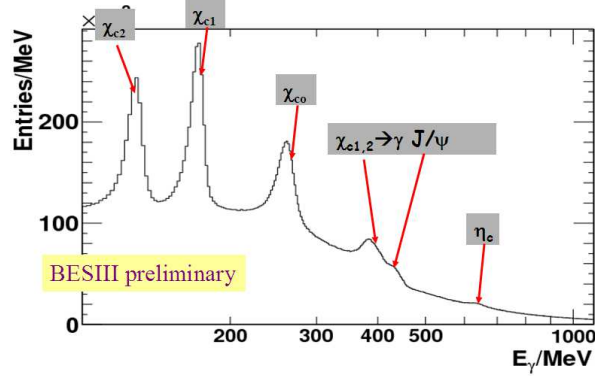}
\caption{Measured inclusive photon spectrum from $\psi(2S)$
decays.} \label{fig8}
\end{figure}

A first glance of the $\psi(2S)$ data shows that a lot of
resonances can be clearly seen. Fig.~\ref{fig8} shows the
inclusive photon spectrum from the electromagnetic calorimeter.
Signals from the electromagnetic transition between charmoniuum
states can be well identified and they demonstrate the impressive
performance of the CsI(Tl) crystal calorimeter.

\begin{figure}[htbp]
\centering\vspace{-0.5cm}
\includegraphics*[width=55mm]{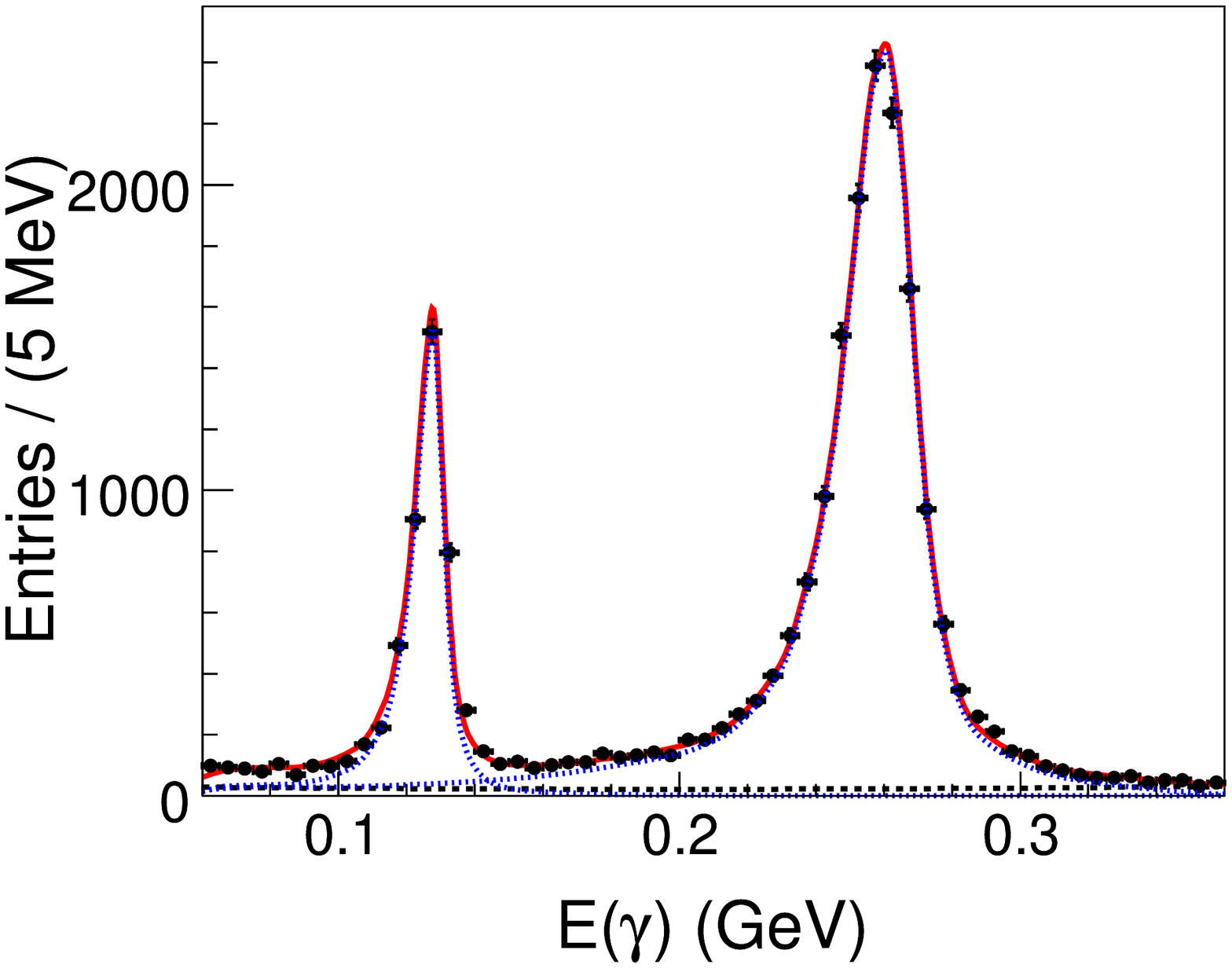}
\includegraphics*[width=55mm]{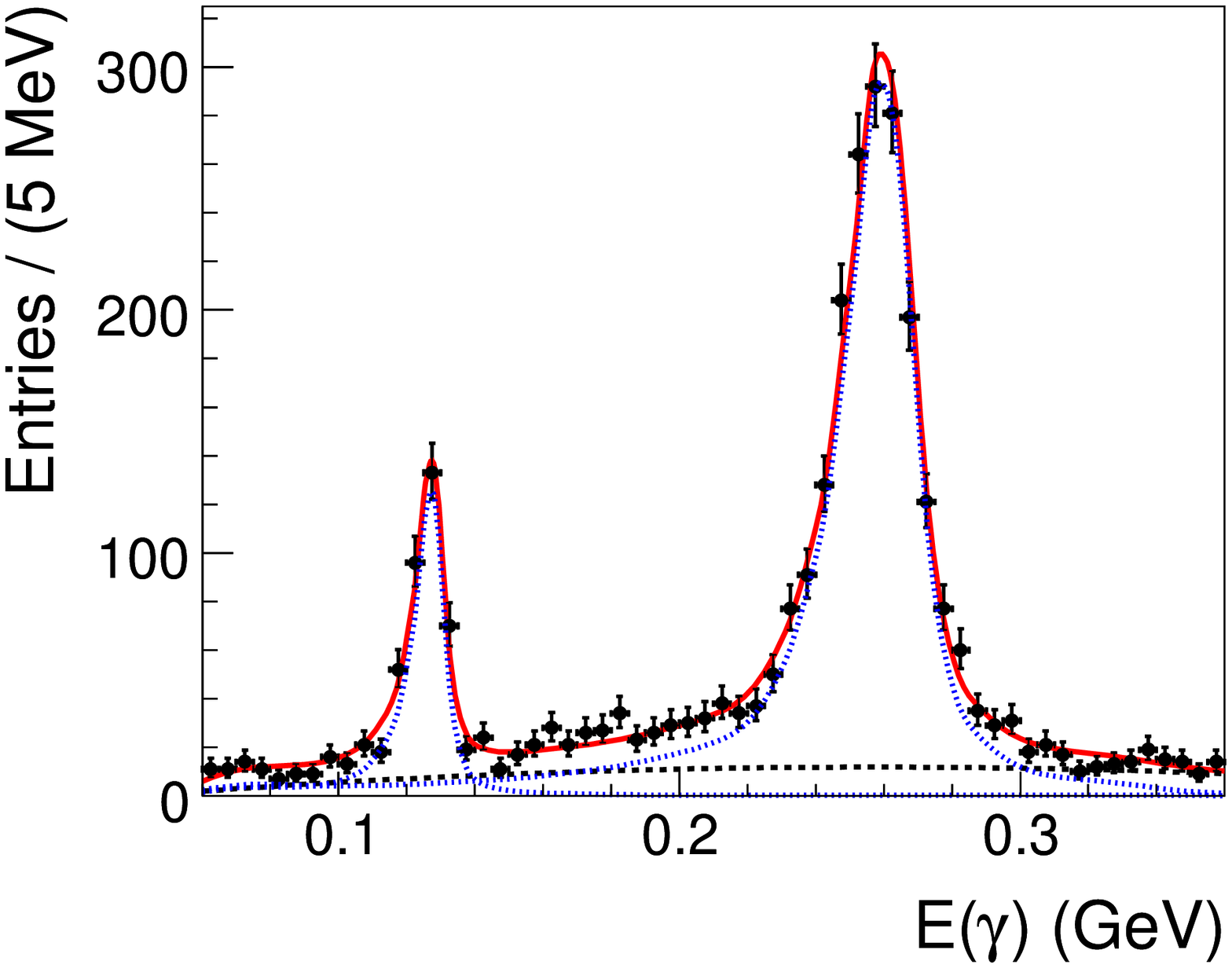}
\caption{Observed $\chi_{c0}$ and $\chi_{c2}$ signal from
$\psi(2S) \rightarrow \gamma\pi^0\pi^0$ (left) and $\psi(2S)
\rightarrow \gamma\eta\eta$ (right) channels. } \label{fig9}
\end{figure}

Initial physics results have been obtained, ranging from the
confirmation of BES-II and CLEO-c results, to completely new
observations. Fig.~\ref{fig9} shows the prompt photon spectrum
from $\psi(2S) \rightarrow\gamma\pi^0\pi^0$ (left) and $\psi(2S)
\rightarrow \gamma\eta\eta$ (right)~\cite{ref:gpipi}. Signals from
$\chi_{c0}$ and $\chi_{c2}$ are observed and their branching
ratios are measured, which are consistent with recent results
from CLEO-c~\cite{ref:CLEOgpp}.

%

\begin{figure}[hbt]
\centering
\includegraphics*[width=60mm]{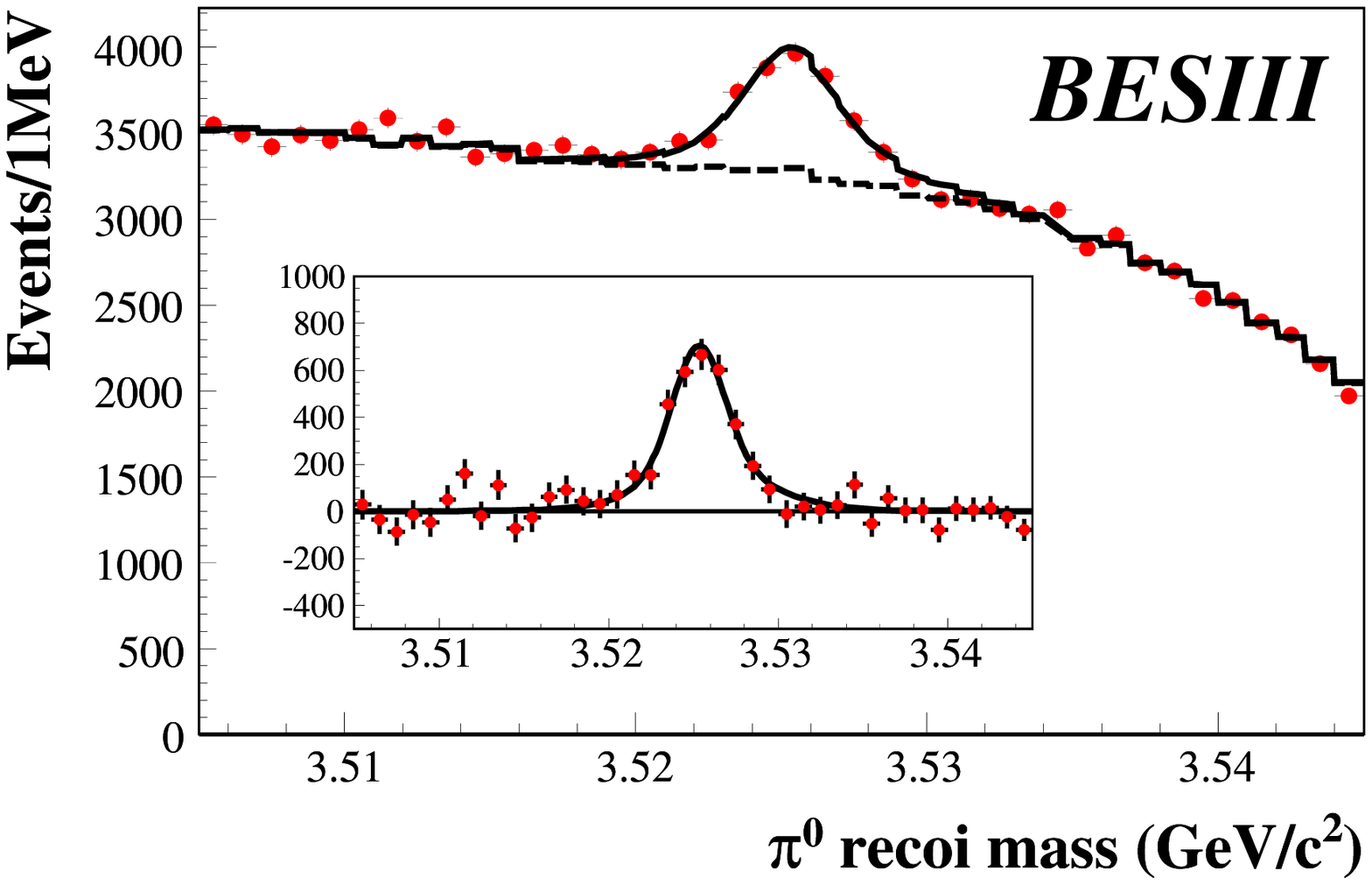 }
\includegraphics*[width=60mm]{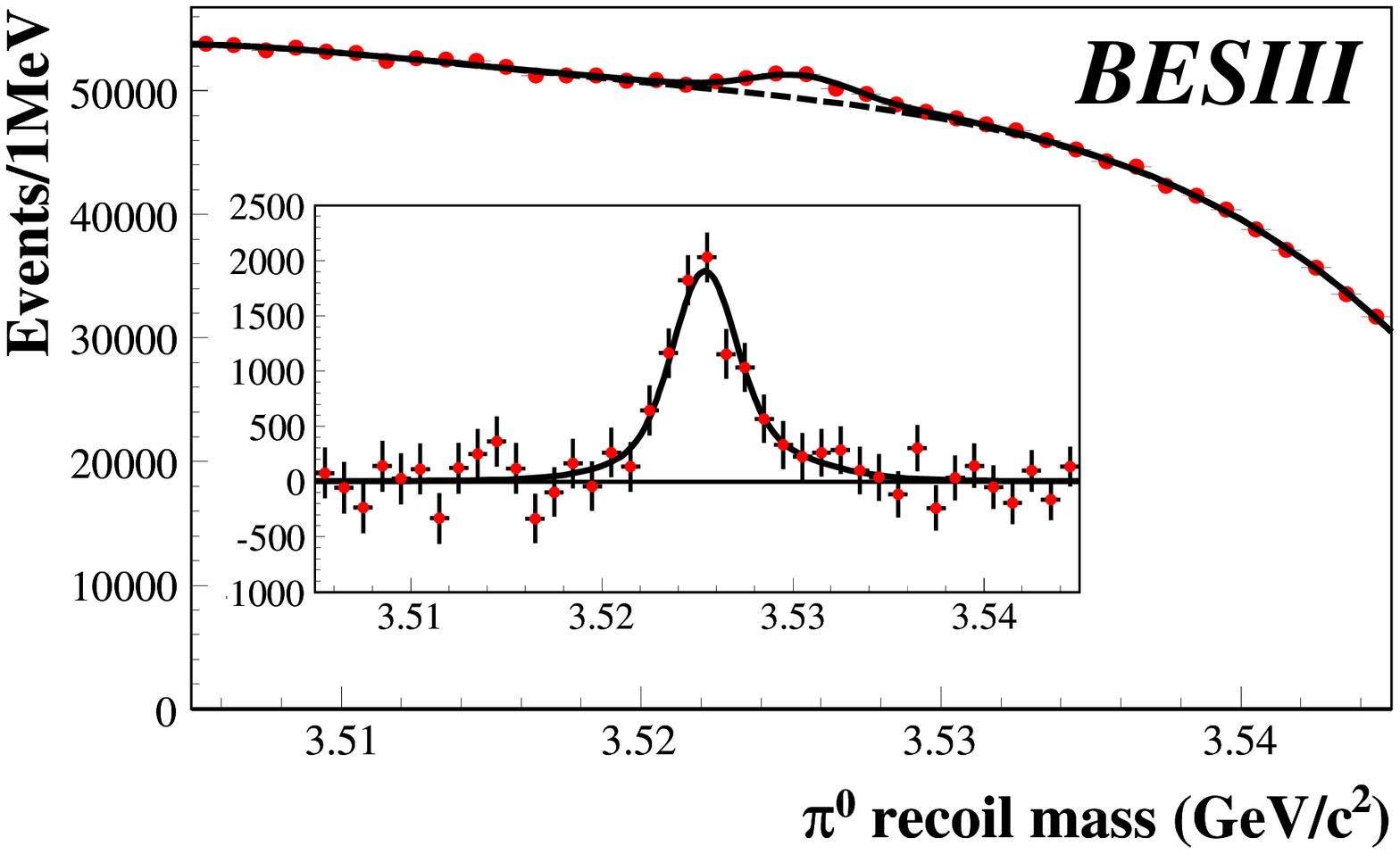 }
\caption{$h_c$ observed in BES-III. Upper: tagging the prompt
photon in the $h_c \rightarrow \eta_c$ decays, lower: tagging
$\pi^0$ from $\psi(2s)\rightarrow \pi^0 h_c$ decays. }
\label{fig10}
\end{figure}

The last member of the charmoniuum family below the open charm
threshold called $h_c$ was observed by CLEO-c in 2005 from
$\psi(2S)$ decays to $\pi^0 h_c$, $h_c\rightarrow \gamma
\eta_c$~\cite{ref:CLEOhc1} and an improved measurement was
performed recently~\cite{ref:CLEOhc}. BESIII performed a similar
analysis with a larger data sample, and a clear signal can be seen
by tagging the prompt photon in the $h_c$ decays~\cite{ref:hc}, as
shown in Fig.~\ref{fig10}. In addition, BESIII tried to look for
inclusive $\pi^0$ from $\psi(2S)$ decays and clear signals can be
also seen. Branching fractions of $\psi(2S) \rightarrow \pi^0
h_c$, $h_c\rightarrow \gamma\eta_c$ can be individually measured
for the first time, together with the width of $h_c$. Results are
listed in the table~\ref{tab:hc} in comparison with recent CLEO
results~\cite{ref:CLEOhc}. Good agreement can be seen.

\begin{table}[htp]
\begin{center}
\footnotesize
\begin{tabular}{|c|c|c|}\hline
Parameters  &  BESIII result   & CLEO results  \\   \hline
$M_{h_c}$  &  $3525.40 \pm 0.13 \pm 0.18$ MeV &  $3525.28\pm 0.19\pm 0.12$ MeV   \\
$\Gamma_{h_c}$  & $(0.73\pm 0.45\pm 0.28) $ MeV &     -  \\
$B(\psi'\rightarrow \pi^0 h_c$ & $(8.4\pm 1.3\pm 1.0)\times 10^{-4}$ & - \\
$B(h_c\rightarrow \gamma\eta_c)$ & $(54.3\pm 6.7\pm 5.2)$\% & - \\
$B(\psi'\rightarrow \pi^0 h_c)\times B(h_c\rightarrow
\gamma\eta_c)$ & $(4.58\pm 0.40 \pm 0.50)\times 10^{-4}$  &
$(4.19\pm 0.32\pm 0.45) \times 10^{-4}$    \\  \hline
\end{tabular}
\caption{Measured results of $h_c$ in comparison with recent CLEO
results. During the fit, $\Gamma_{h_c}$ is floating at BESIII
while CLEO fixes $\Gamma_{h_c}=\Gamma_{\chi_c1}=0.9MeV$}
\label{tab:hc}
\end{center}
\end{table}

Other preliminary results of BESIII include, for example, the
study of $\psi(2S) \rightarrow \gamma VV$, $V=\phi, \omega$,
$\psi(2S)\rightarrow \gamma\gamma V$, $V=\rho,\phi,\omega$,
$\psi(2S) \rightarrow \gamma P$, $P=\pi^0,\eta,\eta^\prime$. New
decay modes have been seen and results will be finalized soon.

\begin{figure}[hbt]
\centering\vspace{-0.5cm}
\includegraphics*[width=55mm]{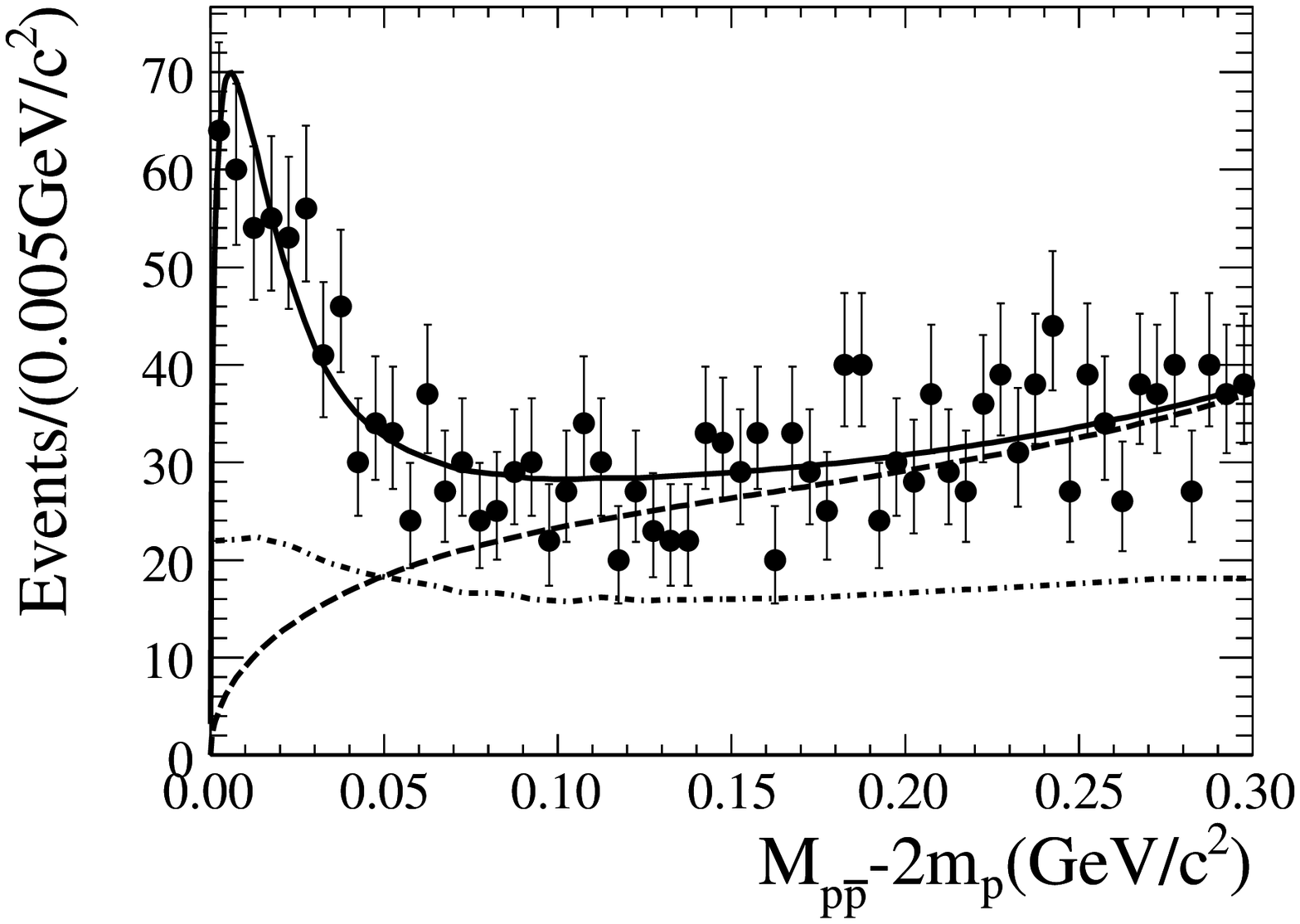}
\includegraphics*[width=60mm]{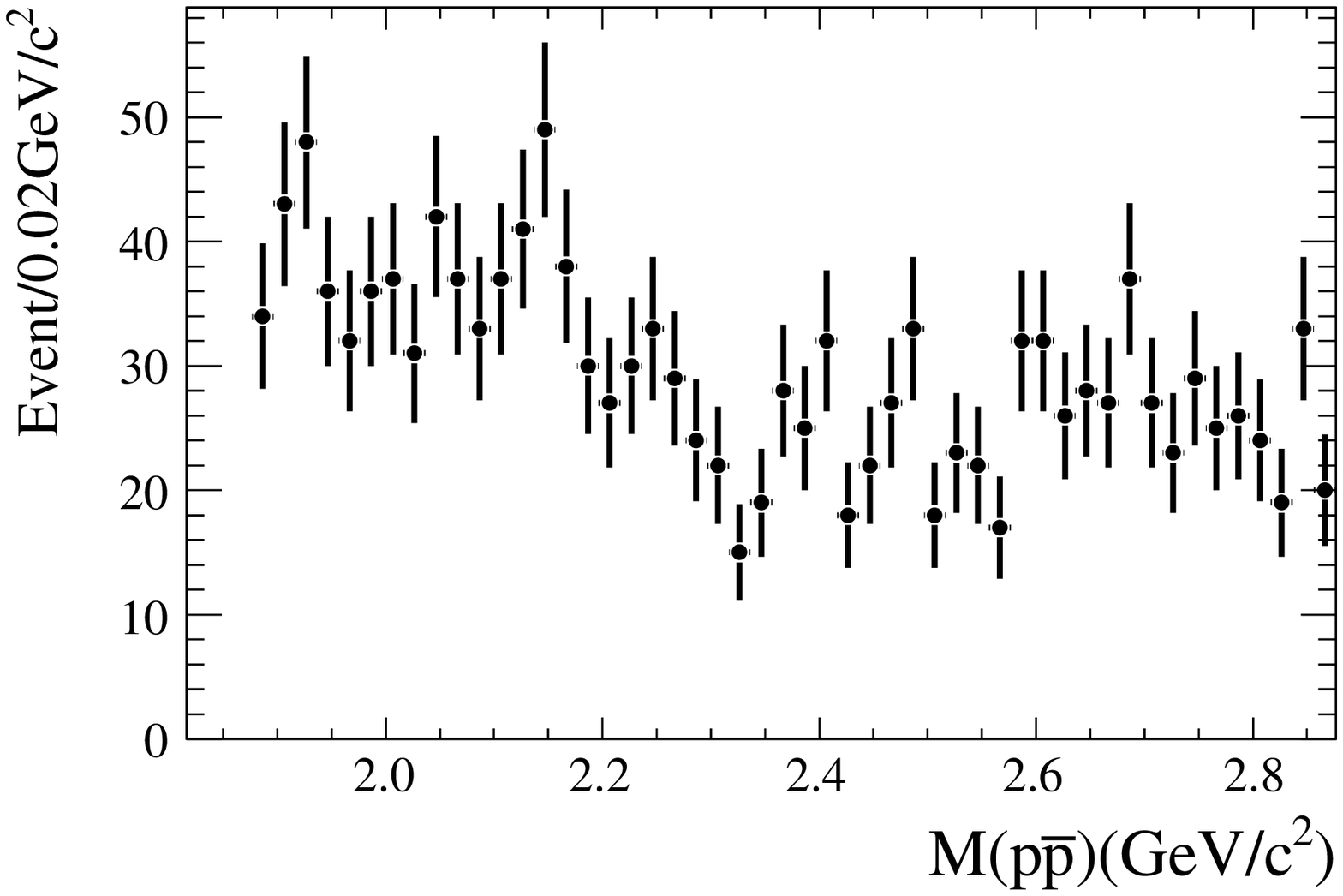}
\caption{Invariant mass of $p\bar{p}$ from (left)
$\psi(2s)\rightarrow \pi\pi J/\psi$, $J\psi \rightarrow \gamma
p\bar{p}$, and (right) $\psi(2s)\rightarrow \gamma p\bar{p}$.}
\label{fig11}
\end{figure}

BES-III also confirmed many observations by
BES-II~\cite{ref:BESII}. Fig.~\ref{fig11} shows the $p\bar{p}$
invariant mass from a) $\psi(2S)\rightarrow \pi\pi J/\psi$,
$J/\psi \rightarrow \gamma p\bar{p}$, and b) $\psi(2S)\rightarrow
\gamma p\bar{p}$~\cite{ref:gpp}. Clearly, a threshold enhancement
can be seen in $J/\psi$ decays, but not in $\psi(2S)$ decays,
consistent with BES-II observations.

\section{Summary}

Charm physics will not stop at BEPCII/BESIII. The newly
operational LHCb experiment, the upgrade of the B-factory at KEK
to be operational in 2014, the PANDA experiment at FAIR planed to
be operational in 2015, will all join the race. The super-flavor
factory planed at FRASCATI and the super-tau-charm factory
proposed at Novosirbisk, may substantial change the field. It is
remarkable that tau-charm collider has a life time much more than
50 years.

\section{Acknowledgments}
I would like to express my sincere thanks to all my BESIII
collaborators and many CLEO people who provides me original
materials. In particular, Haibo Li, helped to prepare this
manuscript, and Ronggang Ping, San Jin, Xiaoyan Shen, Gang Li,
Changzhen Yuan, F. Harris, S. Olsen, R. Poling, D. Asner, P.
Onyisi, M. Shepherd, G. Wilkinson and Zhizhong Xing provided help to
prepare the slides.


\begin{footnotesize}



%

\end{footnotesize}


\end{document}